\documentclass[preprint,prd,showpacs,preprintnumbers,amsmath,amssymb,eqsecnum,nofootinbib]{revtex4}

\usepackage{graphicx}
\usepackage{dcolumn}
\usepackage{bm}

\newcommand{\bea}{\begin{eqnarray}}
\newcommand{\eea}{\end{eqnarray}}
\newcommand{\nn}{\nonumber}

\begin{document}
\title{Target Mass Corrections for the Virtual Photon Structure
Functions to the Next-to-next-to-leading Order in QCD}

\author{Yoshio \surname{Kitadono}}
\email{kitadono@scphys.kyoto-u.ac.jp}
\affiliation{
  Dept. of Physics, Faculty of Science,
  Hiroshima University,
  Higashi Hiroshima 739-8526, Japan
}

\author{Ken \surname{Sasaki}}
\email{sasaki@phys.ynu.ac.jp}
\affiliation{
  Dept. of Physics, Faculty of Engineering,
  Yokohama National University,
  Yokohama 240-8501, Japan
}

\author{Takahiro \surname{Ueda}}
\email{uedat@post.kek.jp}
\affiliation{
  High Energy Accelerator Research Organization (KEK),
  1-1 Oho, Tsukuba, Ibaraki 305-0801, Japan
}

\author{Tsuneo \surname{Uematsu}}
\email{uematsu@scphys.kyoto-u.ac.jp}
\affiliation{
  Dept. of Physics, Graduate School of Science,
  Kyoto University,
  Yoshida, Kyoto 606-8501, Japan
}

\preprint{YNU-HEPTh-07-102}
\preprint{KUNS-2119}


\begin{abstract}
We investigate target mass effects in the unpolarized virtual photon structure
functions $F_2^\gamma(x,Q^2,P^2)$ and $F_L^\gamma(x,Q^2,P^2)$ 
in perturbative QCD for the kinematical 
region $\Lambda^2 \ll P^2 \ll Q^2$, where $-Q^2(-P^2)$  is the mass squared 
of the probe (target) photon and $\Lambda$ is the QCD scale  parameter.
We obtain the Nachtmann moments 
for the structure functions and then, by inverting the moments, we get 
the expressions in closed form for  
$F_2^\gamma(x,Q^2,P^2)$ 
up to the next-to-next-to-leading order  
and for $F_L^\gamma(x,Q^2,P^2)$ up to the next-to-leading order, 
both of which include  the target mass corrections. 
Numerical analysis exhibits that target mass effects appear at large $x$ 
and become sizable near $x_{\rm max}(=1/(1+\frac{P^2}{Q^2}))$, the maximal 
value of  $x$, as the ratio  $P^2/Q^2$ increases.
\end{abstract}

{%
\renewcommand{\baselinestretch}{1.3}\selectfont
\maketitle
}

\section{Introduction}
It is well known that, in $e^+ e^-$ collision experiments, the cross section
for the two-photon processes $e^+ e^- \rightarrow e^+ e^- + {\rm hadrons}$ 
illustrated in Fig. 1 dominates at high energies over other processes such as
one-photon annihilation process $e^+ e^- \rightarrow \gamma^*
\rightarrow {\rm hadrons}$. Here we consider the two-photon processes in the
double-tag events, where both the outgoing $e^+$ and $e^-$ are detected. 
Especially, we investigate the case in which one of the virtual photon
is far off-shell (large $Q^2\equiv -q^2$), while the other is close to
the mass-shell (small $P^2=-p^2$). This process can be viewed as a
deep-inelastic scattering off a photon target \cite{WalshBKT} 
with mass squared $-P^2$,  through which we can study the photon structure 
functions.

In the case of a real photon target ($P^2=0$), unpolarized (spin-averaged)  
photon
structure functions $F_2^\gamma(x,Q^2)$ and $F_L^\gamma(x,Q^2)$ were studied 
first in the
parton model \cite{WalshZerwas}, and then investigated in perturbative 
QCD (pQCD).
In the framework based on the operator product expansion (OPE) 
\cite{CHM} supplemented by the renormalization (RG) group method, 
Witten \cite{Witten} obtained the leading order (LO) QCD contributions to 
$F_2^\gamma$ and $F_L^\gamma$ and, shortly after,  
the next-to-leading order (NLO) QCD corrections to $F_2^\gamma$ were
calculated by Bardeen  and Buras \cite{BB}. 
The same results were rederived by the QCD improved parton model approach 
\cite{Dewitt,GR}. The QCD analysis of the polarized photon structure function 
$g_1^\gamma(x,Q^2)$ 
for the real photon target was performed in the LO \cite{KS} and in the NLO 
\cite{SV,GRS}.

The structure functions $F_2^\gamma(x,Q^2,P^2)$ 
and $F_L^\gamma(x,Q^2,P^2)$ for the case of a virtual photon target ($P^2\ne 0$)
were studied in the  LO~\cite{UW1} and in the
NLO~\cite{UW2} by pQCD. In fact, these structure functions were analyzed
in the kinematical region,
\begin{equation}
\Lambda^2 \ll P^2 \ll Q^2~, \label{Kinematical region}
\end{equation}
where $\Lambda$ is the QCD scale parameter. The advantage of studying a virtual 
photon target in the kinematical region (\ref{Kinematical region}) is that 
we can calculate the whole structure function, its shape and magnitude, 
by the perturbative method. This is contrasted with the case of the real photon target 
where in the NLO there exist nonperturbative pieces.
The virtual photon structure functions $F_2^\gamma$ and $F_L^\gamma$ 
were also studied by using the DGLAP-type QCD
evolution equations \cite{Rossi,DG,GRStratmann,Fontannaz}.
In the same kinematical region (\ref{Kinematical region}), 
the polarized virtual photon structure function $g_1^\gamma(x,Q^2,P^2)$  was
investigated up to the NLO in QCD in Ref.\cite{SU1} and in the second paper of~\cite{GRS}.
Moreover, the polarized parton distributions inside the virtual photon were
analyzed in \cite{SU2}. Recently the first moment of $g_1^\gamma(x,Q^2,P^2)$
was calculated up to the next-to-next-to-leading order (NNLO)~\cite{SUU}.
For more  information on the recent theoretical and experimental
investigation of unpolarized and polarized photon structure, see the review
articles \cite{Krawczyk}.

In our previous paper~\cite{USU}, we have studied 
the unpolarized virtual photon structure functions, 
$F_2^\gamma(x,Q^2,P^2)$ up to the NNLO and $F_L^\gamma(x,Q^2,P^2)$ 
up to the NLO,  in pQCD for the kinematical region (\ref{Kinematical
region}).  This investigation became possible thanks to the recent
three-loop calculations of the  parton-parton as well as photon-parton
splitting functions
\cite{MVV1, MVV2, MVV3}.
There we have considered the
logarithmic corrections arising from the QCD higher-order effects up to the NNLO,
and ignored all the power corrections of the form $(P^2/Q^2)^k\ \  (k=1,2, \cdots)$ 
coming either
from target mass effects  or from higher-twist effects. 

\begin{figure}
\begin{center}
\includegraphics[scale=0.4]{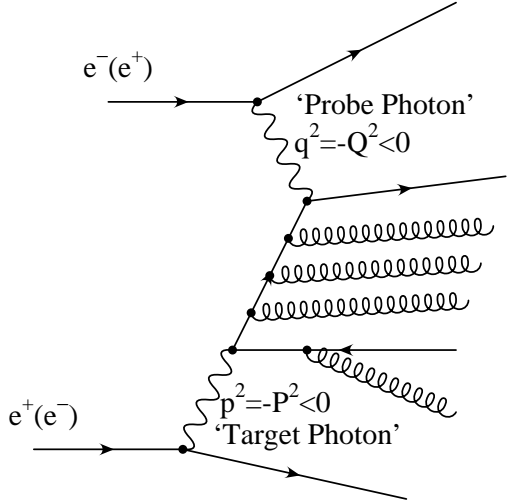}
\vspace{-0.5cm}
\caption{\label{Two Photon Process} Deep inelastic scattering on a 
virtual photon in the $e^+~e^-$
collider experiments.}
\end{center}
\end{figure}
In fact, 
if the target is a real photon ($P^2=0$),  there is no need to consider
target mass corrections. But when the target becomes off-shell, for
example, $P^2 \geq M^2$, where $M$ is the nucleon mass, and for
relatively low values of $Q^2$,  contributions suppressed by powers of
$P^2/Q^2$ may become important.  Then we need to take into account these
target mass contributions just like  the case of the nucleon structure
functions. The consideration of target mass  effects (TME) is important
by another reason.  For the virtual photon target, the maximal value of
the Bjorken variable $x$  is not 1 but 
\begin{equation}
x_{\rm max}=\frac{1}{1+\frac{P^2}{Q^2}}~, \label{xmax}
\end{equation}
due to the constraint $(p+q)^2 \ge 0$, which is in contrast to  
the nucleon case where $ x_{\rm max}= 1$.
The structure functions should 
vanish at $x=x_{\rm max}$. However, both the QCD NNLO  result  for   
$F_2^\gamma(x, Q^2, P^2)$ and the NLO result for
$F_L^\gamma(x,Q^2,P^2)$~\cite{USU} show that  the predicted graphs do not
vanish but remains finite at
$x=x_{\rm max}$. 
This flaw is coming from the fact that TME have not been taken into account
 in the analysis.  
The target mass corrections have been studied in the past for the cases of 
unpolarized~\cite{NACHT,GP,RGP} and polarized~\cite{WAND,MU,KU,PR,BK} nucleon 
structure functions. As for the polarized virtual photon structure functions
$g_1^\gamma (x,Q^2,P^2)$ and $g_2^\gamma (x,Q^2,P^2)$, TME have been studied 
in Ref.~\cite{BSU}. 

In the present paper, we investigate the TME  for the
unpolarized virtual photon structure functions,  
$F_2^\gamma(x,Q^2,P^2)$ up to the NNLO and $F_L^\gamma(x,Q^2,P^2)$ 
up to the NLO, in pQCD. 
We use the framework of the OPE supplemented by the RG method. 
The photon matrix elements of the 
relevant traceless operators in the OPE are expressed by traceless tensors.  
These tensors contain many trace terms so that they satisfy the tracelessness
conditions. The basic idea for computing the 
target mass corrections
is to take account of these trace terms in the traceless tensors properly. 
There are two methods used so far for collecting all those trace terms. 
One, which was introduced by Nachtmann~\cite{NACHT}, is to make use of 
Gegenbauer polynomials to express the contractions between 
$q_{\mu_1}\cdots q_{\mu_{n}}$ and the traceless tensors~\cite{NACHT, WAND,
MU, KU}. This method  leads to the Nachtmann moments for the 
operators  with definite spin. 
The other, first used by Georgi and Politzer~\cite{GP}, is to write  
traceless tensors explicitly and then to collect trace terms and sum them up. 
Through the latter approach, the moments of structure functions are expressed 
as functions of the reduced operator matrix elements and coefficient functions 
with different spins. Actually both methods give equivalent
results.  In this paper we 
apply the former method to study the target mass corrections 
to the structure functions $F_2^\gamma$ and $F_L^\gamma$.

In the next section we discuss the framework for analyzing the TME based 
on the OPE. We introduce  Gegenbauer
polynomials to take account of the trace terms properly. In section 3 we derive
the Nachtmann moments for the structure functions using the orthogonality
relations of  Gegenbauer polynomials.  In section 4, by inverting the
Nachtmann moments, we obtain the explicit expression for  
$F_2^\gamma (x,Q^2,P^2)$ (for $F_L^\gamma (x,Q^2,P^2)$)
evaluated up to the NNLO (up to the NLO) 
with TME included. In section 5 we perform the numerical analysis and
show that target mass corrections become sizable near $x_{\rm max}$. 
The final section is devoted to the conclusion.

\section{Operator Product Expansion\label{Framework}}
We analyze the virtual photon structure functions 
$F_2^{\gamma}(x,Q^2,P^2)$ and $F_L^{\gamma}(x,Q^2,P^2)$ using the
theoretical framework based on the OPE and the RG method. 
Unless otherwise stated, we will follow the notation of Ref.\cite{BB}.
\begin{figure}
\begin{center}
\includegraphics[scale=0.4]{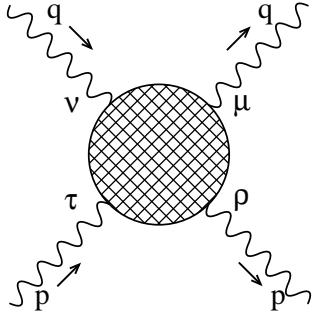}
\vspace{-0.5cm}
\caption{\label{FourPhotons} Forward scattering of a virtual photon with momentum $q$ and 
another virtual photon with momentum $p$. The Lorentz indices are denoted by $\mu, \nu, \rho, \tau$.}
\end{center}
\end{figure}
Let us consider the forward virtual photon scattering amplitude for 
$\gamma(q)+\gamma(p)\rightarrow \gamma(q)+\gamma(p)$ illustrated in Fig.\ref{FourPhotons},
\begin{equation}
T_{\mu\nu\rho\tau}(p,q)=i\int d^4 x d^4 y d^4 z e^{iq\cdot x}e^{ip\cdot (y-z)}
\langle 0|T(J_\mu(x) J_\nu(0) J_\rho(y) J_\tau(z))|0\rangle~,
\end{equation}
where $J_\mu$ is the electromagnetic current.
Its absorptive part is related to the structure tensor 
$W_{\mu\nu\rho\tau}(p,q)$ for the target photon with mass squared 
$p^2=-P^2$ probed by the photon with $q^2=-Q^2$:
\begin{equation}
W_{\mu\nu\rho\tau}(p,q)=\frac{1}{\pi}{\rm Im}T_{\mu\nu\rho\tau}(p,q)~.
\label{StructureTensor}
\end{equation}
Taking a spin average for the target photon, we get
\bea
W_{\mu\nu}^\gamma(p,q)&=&\frac{1}{2}\sum_\lambda\epsilon^{\rho*}_{(\lambda)}(p)
W_{\mu\nu\rho\tau}(p,q)\epsilon^{\tau}_{(\lambda)}(p)~\nonumber\\
&=&-\frac{1}{2}g^{\rho\tau}W_{\mu\nu\rho\tau}(p,q)
=\frac{1}{2}\int d^4 x e^{iqx}\langle \gamma (p)|J_\mu(x) J_\nu(0)|\gamma (p)\rangle_{\rm spin\  av}.
\label{Wmunu}
\eea
Now $W_{\mu\nu}^\gamma(p,q)$ is expressed in terms of two independent structure functions 
$F_L^\gamma(x,Q^2,P^2)$ and $F_2^\gamma(x,Q^2,P^2)$ 
without neglecting the target mass squared $p^2$
(see Appendix~\ref{StructureFunction}):
\bea
W_{\mu\nu}^\gamma(p,q)&=&e_{\mu\nu}\left\{\frac{1}{x}
F_L^\gamma+\frac{p^2q^2}{(p\cdot q)^2}\frac{1}{x}F_2^\gamma
\right\}
+d_{\mu\nu}\frac{1}{x}F_2^\gamma~, 
\label{DefF2FL}
\eea
where
\bea
&&e_{\mu\nu}=g_{\mu\nu}-\frac{q_\mu q_\nu}{q^2} \label{emunu}\\
&&d_{\mu\nu}=-g_{\mu\nu}+\frac{p_\mu q_\nu +p_\nu q_\mu}{p\cdot q}-\frac{
p_\mu p_\nu}{(p\cdot q)^2}q^2~, \label{dmunu}
\eea
and $x$ is the Bjorken variable defined 
by $x=Q^2/2p\cdot q$. 

Applying OPE for the product of two electromagnetic currents at short distance we get
\bea
i\int d^4x e^{iqx}T(J_\mu(x)J_\nu(0))&=&
\left(g_{\mu\nu}-\frac{q_\mu q_\nu}{q^2}  \right)
\sum_{n=0\atop n={\rm even}}\left(\frac{2}{Q^2}\right)^n \hspace{-0.3cm}
q_{\mu_1}\cdots q_{\mu_{n}}
\sum_i C^i_{L,n} O_i^{\mu_1\cdots \mu_{n}} \nonumber\\
&+& \Bigl(-g_{\mu\lambda} g_{\nu\sigma}q^2+g_{\mu\lambda}q_\nu q_\sigma
+g_{\nu\sigma}q_\mu q_\lambda-g_{\mu\nu}q_\lambda q_\sigma \Bigr) \nonumber \\
&\times& \sum_{n=2 \atop n={\rm even}} 
\left(\frac{2}{Q^2}\right)^n \hspace{-0.3cm}
q_{\mu_1}\cdots q_{\mu_{n-2}}
\sum_i C^i_{2,n} O_i^{\lambda\sigma\mu_1\cdots \mu_{{n-2}}}
+ \cdots~, \label{fourier-ope}
\eea
where $C^i_{L,n}$ and $C^i_{2,n}$ are the coefficient functions 
which contribute to the structure functions $F_L^\gamma$ and $F_2^\gamma$, 
respectively, and   
$O_i^{\mu_1\cdots \mu_{n}}$ and $O_i^{\lambda\sigma\mu_1\cdots \mu_{{n-2}}}$ are 
spin-$n$ twist-2 operators  (hereafter we often refer 
to $O_i^{\mu_1\cdots \mu_{n}}$ as $O_i^n$). 
The sum on $i$ runs over  the possible twist-2 operators and 
$\cdots$ represents other terms with irrelevant coefficient functions and operators. 
In fact, the
relevant  $O_i^n$ are flavor singlet quark ($\psi$), gluon ($G$), 
flavor nonsinglet quark ($NS$) and photon ($\gamma$) operators.
It is noted that the operators $O_i^n$ are traceless and have totally symmetric 
Lorentz indices $\mu_1\cdots \mu_{n}$ ($\lambda\sigma\mu_1\cdots \mu_{n-2}$). 

The spin-averaged matrix elements of these operators sandwiched 
by the photon states with momentum $p$ are expressed as 
\begin{eqnarray}
\langle \gamma (p)|O_i^{\mu_1\cdots \mu_{n}}|\gamma (p)
\rangle_{\rm spin\  av}&=&
A_n^{i}(\mu^2, P^2)~\{p^{\mu_1}\cdots p^{\mu_n}~-\ {\rm trace\ terms} \}
\nonumber\\
&\equiv&A_n^i(\mu^2,P^2)\{p^{\mu_1}
\cdots p^{\mu_n}\}_n~, 
\end{eqnarray}
where $i=\psi,G,NS,\gamma$, and $A_n^{i}(\mu^2, P^2)$ is the reduced
photon matrix element with
$\mu$ being the renormalization point which we choose  at $\mu^2=P^2$. For
$-p^2=P^2\gg
\Lambda^2$, we can calculate $A_n^i(P^2)$ perturbatively.  The
$\{p^{\mu_1}\cdots p^{\mu_n}\}_n$ denotes the totally  symmetric rank-$n$
tensor formed with the momentum $p$ alone and satisfies the traceless
condition~
$g_{\mu_i \mu_j}\{p^{\mu_1}\cdots p^{\mu_n}\}_n\!=\!0$~.
Taking the spin-averaged photon matrix elements of (\ref{fourier-ope}) 
we obtain for the photon-photon forward-scattering amplitude
\bea
T_{\mu\nu}^\gamma(p,q)&=&i\int d^4x e^{iq\cdot x}
\langle\gamma(p)|T(J_\mu(x)J_\nu(0))|\gamma(p)\rangle_{\rm spin\ av}\nonumber\\
&=&\left(g_{\mu\nu}-\frac{q_\mu q_\nu}{q^2}  \right)
\sum_{n=0\atop n={\rm even}}\left(\frac{2}{Q^2}\right)^n \hspace{-0.3cm}
q_{\mu_1}\cdots q_{\mu_{n}}\{p^{\mu_1}
\cdots p^{\mu_n}\}_n
~\sum_i C^i_{L,n} A_n^i(P^2) \nonumber\\
&+& \Bigl(-g_{\mu\lambda} g_{\nu\sigma}q^2+g_{\mu\lambda}q_\nu q_\sigma
+g_{\nu\sigma}q_\mu q_\lambda-g_{\mu\nu}q_\lambda q_\sigma \Bigr) \nonumber \\
&\times& \sum_{n=2 \atop n={\rm even}} 
\left(\frac{2}{Q^2}\right)^n \hspace{-0.3cm}
q_{\mu_1}\cdots q_{\mu_{n-2}}\{p^\lambda p^\sigma  p^{\mu_1} \cdots p^{\mu_{n-2}}\}_n
\sum_i C^i_{2,n} A_n^i(P^2)~. \label{forward-amp}
\eea
The basic idea for treating  target mass corrections exactly is 
to take account of trace terms in the traceless tensors properly.
We evaluate the contraction between $q_{\mu_1}\cdots q_{\mu_n}$ and the traceless 
tensors without neglecting any of the trace terms.  The results are expressed in terms of
Gegenbauer polynomials~\cite{NACHT, WAND, MU}:
\bea
&&q_{\mu_1}\cdots q_{\mu_n}\{p^{\mu_1}
\cdots p^{\mu_n}\}_n=a^nC_n^{(1)}(\eta)\label{contraction-1}~,\\
&&q_{\mu_1}\cdots q_{\mu_{n-2}}\{p^\lambda p^\sigma p^{\mu_1}
\cdots p^{\mu_{n-2}}\}_n=\frac{1}{n(n-1)}
\left[
\frac{g^{\lambda\sigma}}{Q^2}a^n 2 C_{n-2}^{(2)}(\eta)
+\frac{q^\lambda q^\sigma}{Q^4}a^n 8C_{n-4}^{(3)}(\eta)\right.\nonumber\\
&&\hspace{5.0cm}\left.+p^\lambda p^\sigma a^{n-2} 2 C_{n-2}^{(3)}(\eta)
+\frac{p^\lambda q^\sigma+q^\lambda p^\sigma}{Q^2}
a^{n-1} 4 C_{n-3}^{(3)}(\eta)
\right],\label{contraction-2}
\eea
where 
\bea
a=-\frac{1}{2}PQ, \quad
\eta=-\frac{p\cdot q}{PQ}~,\label{a-eta}
\eea
and $C_n^{(\nu)}(\eta)$'s are Gegenbauer polynomials (see
Appendix~\ref{Gegenbauer}).  Recall that in the case of a nucleon target with mass
$M$, we had $p^2=M^2$,  
$\eta=ip\cdot q/MQ$ and $a=-\frac{1}{2}iMQ$ \cite{WAND}. 
In the photon case, we have $p^2=-P^2$ instead, and thus,   
replacing $M$ with $-iP$,  we obtain the expressions for $a$ and $\eta$ in 
(\ref{a-eta}). 
The derivation of Eqs.(\ref{contraction-1}) and (\ref{contraction-2})
are given in Appendix~\ref{Contraction}.

We decompose the amplitude $T_{\mu\nu}^\gamma(p,q)$ as
\begin{equation}
T_{\mu\nu}^\gamma(p,q)=e_{\mu\nu}\left\{\frac{1}{x}
T_L^\gamma+\frac{p^2q^2}{(p\cdot q)^2}\frac{1}{x}(p\cdot
q)T_2^\gamma
\right\}
+d_{\mu\nu}\frac{1}{x}(p\cdot q) T_2^\gamma~,
\end{equation}
then, using the results (\ref{contraction-1}) and (\ref{contraction-2}),
we find from Eq.(\ref{forward-amp})
\bea
&&\frac{1}{x}(p\cdot q) T_2^\gamma=
\sum_{n=2\atop n={\rm even}}\left(-\frac{P}{Q}\right)^n 
8\eta^2 C_{n-2}^{(3)}(\eta)\frac{1}{n(n-1)}~M_{2,n}^\gamma~,
\label{T2-amp}\\
&&\frac{1}{x}
T_L^\gamma+\frac{p^2q^2}{(p\cdot q)^2}\frac{1}{x}(p\cdot
q)T_2^\gamma=
\sum_{n=0\atop n={\rm even}}\left(-\frac{P}{Q}\right)^n 
C_n^{(1)}(\eta)~M_{L,n}^\gamma\nonumber\\
&&\qquad+
\sum_{n=2\atop n={\rm even}}\left(-\frac{P}{Q}\right)^n 
\frac{4}{n(n-1)}\times
\left[
C_{n-2}^{(2)}(\eta)-2C_{n-4}^{(3)}(\eta)+4\eta C_{n-3}^{(3)}(\eta)
\right] M_{2,n}^\gamma
~,\label{TL-amp}
\eea
where we have defined
\bea
M_{2,n}^\gamma\equiv \frac{1}{2}\sum_iC^i_{2,n}(Q^2,P^2,g)A_n^i(P^2),\quad
M_{L,n}^\gamma\equiv \frac{1}{2}\sum_iC^i_{L,n}(Q^2,P^2,g)A_n^i(P^2)~.
\eea

\section{Nachtmann Moments \label{NachtmannMoments}}

We derive the Nachtmann moments for the definite spin-$n$ contributions, $M_{2,n}^\gamma$
and $M_{L,n}^\gamma$. First we write the dispersion relations for 
$T_2^\gamma$ and $T_L^\gamma$, and we denote
\bea
F_2^\gamma=\frac{1}{\pi}{\rm Im}\ (p\cdot q) T_2^\gamma, \qquad
F_L^\gamma=\frac{1}{\pi}{\rm Im} T_L^\gamma~.
\eea
Then, using the orthogonality relation (\ref{orthogonal}) and the integration formula 
 (\ref{IntegralGegenbauer}) for the Gegenbauer polynomials $C_n^{(\nu)}(\eta)$, we 
can project
out $M_{2,n}^\gamma$ and $M_{L,n}^\gamma$.
The results are as follows:
\bea
\mu_{2,n}^\gamma(Q^2,P^2)
&\equiv&
\int_0^{x_{\rm max}}dx\frac{1}{x^3}\xi^{n+1}
\left[
\frac{3+3(n+1)r+n(n+2)r^2}{(n+2)(n+3)}\right]
F_2^\gamma(x,Q^2,P^2)=M_{2,n}^\gamma, \label{NachtmannM_2}\nonumber\\ 
&&\\
\mu_{L,n}^\gamma(Q^2,P^2)&\equiv&
\int_0^{x_{\rm max}} dx 
\frac{1}{x^3}\xi^{n+1}\left[F_L^\gamma(x,Q^2,P^2)\right.\nonumber\\
&&\hspace{1.2cm}+\left.\frac{4P^2 x^2}{Q^2}
\frac{(n+3)-(n+1)\xi^2 P^2/Q^2}{(n+2)(n+3)}F_2^\gamma(x,Q^2,P^2)\right]
=M_{L,n}^\gamma.
\label{NachtmannM_L}
\eea
The Nachtmann moments $\mu_{2,n}^\gamma$ and $\mu_{L,n}^\gamma$ are given by
the weighted integrals of the structure functions $F_2^\gamma$ and $F_L^\gamma$
and are equal to the  definite spin-$n$ contributions, $M_{2,n}^\gamma$
and $M_{L,n}^\gamma$, respectively.
The variables $r$ and $\xi$ are defined as
\bea
r\equiv \sqrt{1-\frac{4P^2x^2}{Q^2}}\label{xi}~,\qquad 
\xi\equiv\frac{2x}{1+\sqrt{1-\frac{4P^2x^2}{Q^2}}}=\frac{2x}{1+r}.
\eea
We see from Eq.(\ref{xmax}) that the maximal value of $x$ is not 1 but $1/[1+(P^2/Q^2)]$. 
Therefore, 
the allowed ranges of $r$ and $\xi$ turn out to be $r_{\rm min}\leq
r\leq 1$ and  $0\leq\xi\leq 1$, respectively, where $r_{\rm
min}=r(x_{\rm max})=(1-P^2/Q^2)/(1+P^2/Q^2)$ and $\xi(x_{\rm max})=1$.

We now outline how to derive the Nachtmann moments for the case of 
$F_2^\gamma(x,Q^2,P^2)$ given in
(\ref{NachtmannM_2}). Since $(p\cdot q)/x=(-PQ\eta)(-P/Q)2\eta$, we see  
that $T_2^\gamma$ in (\ref{T2-amp})  is expressed as
\bea
 T_2^\gamma(x,Q^2,P^2)=\frac{1}{Q^2}\sum_{n=2\atop n={\rm even}}
\left(-\frac{P}{Q}\right)^{n-2}\cdot 8C_{n-2}^{(3)}(\eta)
\frac{1}{n(n-1)}M_{2,n}^\gamma~.\label{T2}
\eea
By the use of orthogonality relation of the Gegenbauer polynomials
(\ref{orthogonal}) for $\nu=3$ we get
\bea
\int_{-1}^{1}(1-\eta^2)^{5/2}C_{n-2}^{(3)}(\eta)~T_2^\gamma d\eta
=\frac{1}{Q^2}\frac{\pi}{2^4}\left(-\frac{P}{Q}\right)^{n-2}
(n+2)(n+3)M_{2,n}^\gamma~.\label{LHS0}
\eea
Applying the dispersion relation, we can relate the full amplitude with its
absorptive part:
\bea
T_2^\gamma(\omega)
=\int_{\omega_{\rm min}}^\infty\left(
\frac{1}{\omega'-\omega}+\frac{1}{\omega'+\omega}
\right) W_2^\gamma(\omega')d\omega'
~,\quad \omega_{\rm min}=\frac{1}{x_{\rm min}}
\eea
where $\omega=2p\cdot q/Q^2=1/x$ 
and denoting $\zeta=(\omega'Q)/(-2P)$, 
$\eta=(\omega Q)/(-2P)$ we derive
\bea
&&\hspace{-0.5cm}\int_{-1}^1(1-\eta^2)^{5/2}C_{n-2}^{(3)}(\eta)T_2^\gamma d\eta
=2\int_{\omega_{\rm min}}^\infty d\omega' W_2^\gamma(\omega',Q^2,P^2)\frac{Q}{(-2P)}
\int_{-1}^1 \frac{(-1)}{(-\zeta)-\eta}(1-\eta^2)^{5/2}C_{n-2}^{(3)}(\eta)d\eta~,\nonumber\\
\eea
where we have noted $C_{n-2}^{(3)}(-\eta)=C_{n-2}^{(3)}(\eta)$ for even $n$.
From (\ref{IntegralGegenbauer}) for $m=0$, $\nu=3$, we get
\bea
&&\int_{-1}^1(1-\eta^2)^{5/2}C_{n-2}^{(3)}(\eta)T_2^\gamma d\eta\nonumber\\
&&=\left(\frac{-Q}{2P}\right)\cdot\frac{\pi}{4}\cdot 
\int_{\omega_{\rm min}}^\infty d\omega W_2^\gamma(\omega,Q^2,P^2)
\left(\zeta^2-1\right)
(-1)\left[-\zeta-(\zeta^2-1)^{1/2}\right]^{n+1}
(n+2)(n+3)\nonumber\\
&&\hspace{3cm} \times\left[
1-\frac{6}{n+2}z+\frac{12}{(n+2)(n+3)}z^2\right]~,\label{RHS0}
\eea
where we used the following relation for the hypergeometric 
function with $\zeta=-p\cdot q/PQ$,
\bea 
F(3,-2;n+2;z)=1-\frac{6}{n+2}z+\frac{12}{(n+2)(n+3)}z^2,\quad
z\equiv \frac{-(-\zeta)+(\zeta^2-1)^{1/2}}{2(\zeta^2-1)^{1/2}}~,
\eea
Setting (\ref{LHS0}) equal to (\ref{RHS0}) and changing integration variable
from $\omega$ to $x$, we get
\bea
\int_0^{x_{\rm max}}dx\frac{1}{x^3}\xi^{n+1}\left(1-\frac{4P^2x^2}{Q^2}\right)
\left[1-\frac{6}{n+2}z+\frac{12}{(n+2)(n+3)}z^2\right]F_2^\gamma(x,Q^2,P^2)
=M_{2,n}^\gamma~.\nonumber\\
\eea
Here one should note that $n$ is even and the following relations hold:
\bea
-\zeta=\frac{Q}{2Px},\quad
-\zeta-(\zeta^2-1)^{1/2}=\frac{P}{Q}\xi,\quad z=-\frac{P^2}{Q^2}
\frac{\xi x}{r},\quad \xi=\frac{Q^2}{2P^2x}(1-r)~,
\eea
and so we finally get for $F_2^\gamma(x,Q^2,P^2)$ as
\bea
\int_0^{x_{\rm max}}dx\frac{1}{x^3}\xi^{n+1}
\left[
\frac{3+3(n+1)r+n(n+2)r^2}{(n+2)(n+3)}\right]
F_2^\gamma(x,Q^2,P^2)=M_{2,n}^\gamma(Q^2,P^2)~,
\eea
where the left-hand side is $\mu_{2,n}^\gamma(Q^2,P^2)$, the Nachtmann moment,
which is equal to the definite spin-$n$ contribution $M_{2,n}^\gamma$,
and this is consistent with the previous result for the case of nucleon target
\cite{NACHT,Simula,SteffensMelnitchouk,Schi} with a replacement of 
the variable $M \rightarrow -iP$.

For the longitudinal structure function, we first solve for $\frac{1}{x}T_L^\gamma$ 
from Eqs.(\ref{T2-amp}) and (\ref{TL-amp}) and we get 
(see Appendix~\ref{Gegenbauer})
\bea
\frac{1}{x}T_L^\gamma&=&\sum_{n=0\atop n={\rm even}}\left(-\frac{P}{Q}\right)^n 
C_n^{(1)}(\eta)~M_{L,n}^\gamma\nn\\
&&\quad +\sum_{n=2\atop n={\rm even}}\left(-\frac{P}{Q}\right)^n 
\frac{4}{n(n-1)}\left[
\frac{-2}{n}C_{n-2}^{(3)}(\eta)+\frac{2}{n}C_{n-4}^{(3)}(\eta)
\right] M_{2,n}^\gamma~.\label{tl}
\label{tlamp0}
\eea

Then using the recursion relation (\ref{Recusion3term}) for the case $\nu=1$:
\bea
C_n^{(1)}(\eta)=
\frac{2}{(n+1)(n+2)}C_n^{(3)}(\eta)
-\frac{4}{n(n+2)}C_{n-2}^{(3)}(\eta)+\frac{2}{n(n+1)}C_{n-4}^{(3)}(\eta)
\label{C1C3}~,
\eea
and the orthogonality relation of the Gegenbauer polynomials $C_n^{(3)}$'s
we can derive the recursive relations for the sequence $M_{L,n}^\gamma$'s which
can be solved 
as (See Appendix \ref{FLNachtmann}):
\bea
&&
\int_0^{x_{\rm max}} dx 
\frac{1}{x^3}\xi^{n+1}\left[F_L^\gamma(x,Q^2,P^2)
+\frac{4P^2 x^2}{Q^2}
\frac{(n+3)-(n+1)\xi^2 P^2/Q^2}{(n+2)(n+3)}F_2^\gamma(x,Q^2,P^2)\right]
=M_{L,n}^\gamma~,\nonumber\\
\eea
where the left-hand side integral is the Nachtmann moment $\mu_{L,n}^\gamma$ 
for the longitudinal part. This coincides with the result obtained in 
\cite{NACHT,Simula} after the replacement mentioned above.

\section{Inverting the Nachtmann Moments \label{InvertingNachtmannMoments}}

We now invert the Nachtmann moments to express the structure
functions $F_2^\gamma$ and $F_L^\gamma$ explicitly as functions of $x$, $Q^2$
and $P^2$. We first consider $F_2^\gamma$.
By changing the integration variable from $x$ to $\xi$, we can rewrite
the Nachtmann moments, given in Eqs.(\ref{NachtmannM_2}) and (\ref{NachtmannM_L}), 
as follows:
\bea
M_{2,n}^\gamma&=&\int_0^1d\xi\ \xi^{n-2}(1-\kappa\xi^2)(1+\kappa\xi^2)\nn \\
&&\hspace{-1.5cm}\times
\biggl[\frac{3}{(n+2)(n+3)}+\frac{3(n+1)}{(n+2)(n+3)}\frac{1-\kappa\xi^2}
{1+\kappa\xi^2}
+\frac{n}{n+3}\Bigl( \frac{1-\kappa\xi^2}{1+\kappa\xi^2}  \Bigr)^2
\biggr]~F_2^\gamma(x,Q^2,P^2)\\ \label{NachtmannF_2-1}
M_{L,n}^\gamma&=&\int_0^1d\xi\ \xi^{n-2}(1-\kappa\xi^2)(1+\kappa\xi^2)\nn \\
&&\hspace{-1.5cm}\times
\biggl[F_L^\gamma(x,Q^2,P^2)+4\kappa\frac{\xi^2}{(1+\kappa\xi^2)^2}
\biggl(\frac{1}{n+2}-\kappa \xi^2\frac{n+1}{(n+2)(n+3)}
\biggr)F_2^\gamma(x,Q^2,P^2)
\biggr]~.\label{NachtmannFL-1}
\eea
where we have made use of the following relations:
\bea
x=\frac{\xi}{1+\kappa\xi^2}~, \quad r=\frac{1-\kappa\xi^2}{1+\kappa\xi^2}~,
\quad
\frac{dx}{d\xi}&=&\frac{1-\kappa\xi^2}{(1+\kappa\xi^2)^2}~,
\eea
with $\kappa=P^2/Q^2$. Now we define
\bea
A(\xi)\equiv \frac{1-\kappa\xi^2}{1+\kappa\xi^2}~F_2^\gamma(x,Q^2,P^2)~,
\quad 
B(\xi)\equiv (1-\kappa\xi^2)(1+\kappa\xi^2)F_L^\gamma(x,Q^2,P^2)~.\label{AxiBxi}
\eea
Then the above two moments are written as 
\bea
\frac{M_{2,n}^\gamma}{n(n-1)}&=&
\int_0^1d\xi\ \xi^{n-2}\biggl\{\frac{A(\xi)}{n(n-1)}
-\frac{2\kappa\xi^2~A(\xi)}{n(n+2)}
+\frac{\kappa^2\xi^4~A(\xi)}{(n+2)(n+3)}\biggr\} \\
\frac{M_{L,n}^\gamma}{n+1}&=&
\int_0^1d\xi\ \xi^{n-2}\biggl\{\frac{B(\xi)}{n+1}
+\frac{4\kappa\xi^2~A(\xi)}{(n+1)(n+2)}
-\frac{4\kappa^2\xi^4~A(\xi)}{(n+2)(n+3)}\biggr\}
\eea

The boundary conditions for $A(\xi)$ and $B(\xi)$ are
$A(\xi=1)=B(\xi=1)=0$, since $F_2^\gamma(x_{\rm max},Q^2,P^2)=F_L^\gamma(x_{\rm
max},Q^2,P^2)=0$ and  $\xi(x_{\rm max})=1$. 
Now introducing the following four functions,
\bea
A_0(\xi)=\int^1_{\xi}d\xi'~A(\xi'), \quad
A_{-1}(\xi)&\equiv&\int^1_{\xi}d\xi'~\frac{A(\xi')}{\xi'}, \quad 
A_{-2}(\xi)=\int^1_{\xi}d\xi'~\frac{A(\xi')}{{\xi'}^2}\\
B_{-3}(\xi)&\equiv&\int^1_{\xi}d\xi'~\frac{B(\xi')}{{\xi'}^3}
\eea
and by partial integration we find that the above two moments are written as
\bea
\frac{M_{2,n}^\gamma}{n(n-1)}
&=&\int_0^1d\xi~\xi^{n-2} (1-\kappa\xi^2) \Bigl\{(1+\kappa\xi^2)A_{-1}(\xi)
-\xi A_{-2}(\xi) -\kappa\xi   A_{0}(\xi) \Bigr\},\\
&&\nn\\
\frac{M_{L,n}^\gamma}{n+1}
&=&\int_0^1d\xi~\xi^{n}\Bigl[B_{-3}(\xi)+4\kappa
\Bigl\{(1+\kappa\xi^2)A_{-1}(\xi)
-\xi A_{-2}(\xi) -\kappa\xi   A_{0}(\xi) \Bigr\}\Bigr].
\eea

Inverting the moments we get
\bea
G(\xi)&=&\frac{1}{2\pi i}\int_{c-i\infty}^{c+i\infty}
dn \,\xi^{-n+1}\biggl\{ \frac{M_{2,n}^\gamma}{n(n-1)} \biggr\} \nn\\
&=& (1-\kappa\xi^2) \Bigl\{(1+\kappa\xi^2)A_{-1}(\xi)
-\xi A_{-2}(\xi) -\kappa\xi   A_{0}(\xi) \Bigr\},
\label{InversionF2}\\ &&\nn\\
S(\xi)&=&\frac{1}{2\pi i}\int_{c-i\infty}^{c+i\infty}
dn \,\xi^{-n-1}\biggl\{ \frac{M_{L,n}^\gamma}{n+1} \biggr\} \nn\\
&=&B_{-3}(\xi)+\frac{4\kappa}{1-\kappa\xi^2}G(\xi),\label{InversionFL}
\eea
where in Eq.(\ref{InversionFL}) we have used the result of Eq.(\ref{InversionF2}).

We further introduce the following functions,
\bea
H(\xi)&\equiv&-\frac{dG(\xi)}{d\xi}=\frac{1}{2\pi i}\int_{c-i\infty}^{c+i\infty}
dn \,\xi^{-n} \frac{M_{2,n}^\gamma}{n} \label{Hxi}\\
F(\xi)&\equiv&-\frac{dH(\xi)}{d\xi}=\frac{1}{2\pi i}\int_{c-i\infty}^{c+i\infty}
dn \,\xi^{-n-1} M_{2,n}^\gamma \label{Fxi}\\
F_L(\xi)&\equiv&-\xi \frac{dS(\xi)}{d\xi}=\frac{1}{2\pi i}\int_{c-i\infty}^{c+i\infty}
dn \,\xi^{-n-1} M_{L,n}^\gamma~.
\label{HF2Fl}
\eea

Differentiating both sides of Eqs.(\ref{InversionF2}-\ref{InversionFL}) 
with respect to $\xi$, we get the relations between ${A_{-1}}(\xi)$,
$({A_{-2}}(\xi)+\kappa{A_0}(\xi))$, $A(\xi)$,
$B(\xi)$ and $G(\xi)$, $H(\xi)$, $F_2(\xi)$ and $F_L(\xi)$.
Now solving for $A(\xi)$ and $B(\xi)$ and, then recalling Eq.(\ref{AxiBxi}), 
we obtain
\bea
F_2^\gamma(x,Q^2,P^2)&=&\frac{x^2}{r^3}F(\xi)-6\kappa\frac{x^3}{r^4}H(\xi)
+12\kappa^2\frac{x^4}{r^5}G(\xi)~,\label{TargetMassF2}\\
F_L^\gamma(x,Q^2,P^2)&=&\frac{x^2}{r}F_L(\xi)-4\kappa\frac{x^3}{r^2}H(\xi)
+8\kappa^2\frac{x^4}{r^3}G(\xi)~.\label{TargetMassFL}
\eea

Eqs.(\ref{TargetMassF2}), (\ref{TargetMassFL}) are the final formulas for 
the photon structure functions $F_2^\gamma$ and $F_L^\gamma$ when  target 
mass effects are taken into account. 
They can also be derived from the method of Georgi and Politzer \cite{GP}.
Once $M_{2,n}^\gamma$ and 
$M_{L,n}^\gamma$ in Eq.(2.16), are given, then we can calculate the four
\lq\lq profile\rq\rq\  functions
$G(\xi)$, $H(\xi)$, $F(\xi)$ and $F_L(\xi)$ through
Eqs.(\ref{InversionF2}-\ref{HF2Fl}), and by using 
Eqs.(\ref{TargetMassF2}-\ref{TargetMassFL}) we can predict whole structure 
functions with target mass corrections. 

Note that in the above expressions, the $Q^2$- as well as $P^2$-
dependence of $F(\xi)$, $G(\xi)$, $H(\xi)$ and $F_L(\xi)$ are implicit, since
they are given by $M_{2,n}^\gamma$ and $M_{L,n}^\gamma$ which depend on
$Q^2$ as well as on $P^2$. 
If we take the $\kappa\rightarrow 0$ limit, the above expression
reduces to that for the $F_2^\gamma$ and $F_L^\gamma$ without TME.
And in the absence of TME we have
\bea
M_{2,n}^\gamma= \int_0^1 dx x^{n-2} 
F_2^\gamma(x,Q^2,P^2),\quad
M_{L,n}^\gamma= \int_0^1 dx x^{n-2} 
F_L^\gamma(x,Q^2,P^2)~.
\eea

From our previous QCD calculation \cite{USU}
of $M_{2,n}^\gamma$ and $M_{L,n}^\gamma$ 
we already know  the three functions, $F(\xi)$, 
$G(\xi)$ and $H(\xi)$ to NNLO and $F_L(\xi)$ to NLO, 
so we can evaluate the photon structure functions 
with TME to the same accuracies as in the case we neglect
TME. 

\section{Numerical Analysis \label{NumericalAnalysis}}
In this section we perform a numerical analysis for the structure
functions $F_2^\gamma$ and $F_L^\gamma$ when TME are included.
We first compute the four profile functions $F$, $G$, $H$ and $F_L$ 
which are given in Eq.(\ref{InversionF2}) and Eqs.(\ref{Hxi})-(\ref{HF2Fl}).
We use the QCD results for $M_{2,n}^\gamma$
and $M_{L,n}^\gamma$, which have been calculated up to the NNLO 
and the NLO in QCD, respectively, in Ref.~\cite{USU}.
Indeed,  the expressions of $M_{2,n}^\gamma$ and $M_{L,n}^\gamma$ are given 
in the right-hand sides of Eq.(2.29) and Eq.(6.3) of Ref.~\cite{USU}.
In Fig. 3 we have plotted the functions $F$, $G$, $H$ and $F_L$ 
as  functions of $x$ for the case of $Q^2=30{\rm GeV}^2$ and
$P^2=1{\rm GeV}^2$ with $x_{\rm max}=0.968$. We take $\Lambda=0.2$ GeV for the 
QCD parameter and $n_f=4$ for the number of active quark flavors throughout 
our numerical analysis. Note that we have multiplied each function by a 
suitable power of $x$ to accommodate four functions in a single graph. 
The Bjorken variable $x$ ranges from $0$ to $x_{\rm max}$.
 
Now inserting the functions $F$, $G$, $H$ and $F_L$ into 
 Eqs. (\ref{TargetMassF2}) and (\ref{TargetMassFL}), we obtain the graphs of 
$F_2^\gamma(x,Q^2,P^2)$ and $F_L^\gamma(x,Q^2,P^2)$ as 
functions of $x$, which are shown in Fig. 4 and Fig. 5, respectively. 
We observe that TME become sizable at larger
$x$ region. While TME enhances $F_2^\gamma$ at larger $x$, it reduces 
$F_L^\gamma$.  In fact, $F_2^\gamma$ becomes maximum at $x$ very close 
to the maximal value of $x$, $x_{\rm max}$ (1) for the case with (without) TME.
The target mass correction is of order 10 $\%$ when compared at the maximal 
values for $F_2^\gamma$. In the case of $F_L^\gamma$, the maximal value is
attained in the middle $x$, where the TME reduces the $F_L^\gamma$ about 5 $\%$.
\begin{figure}
\begin{center}
\includegraphics[scale=1.0]{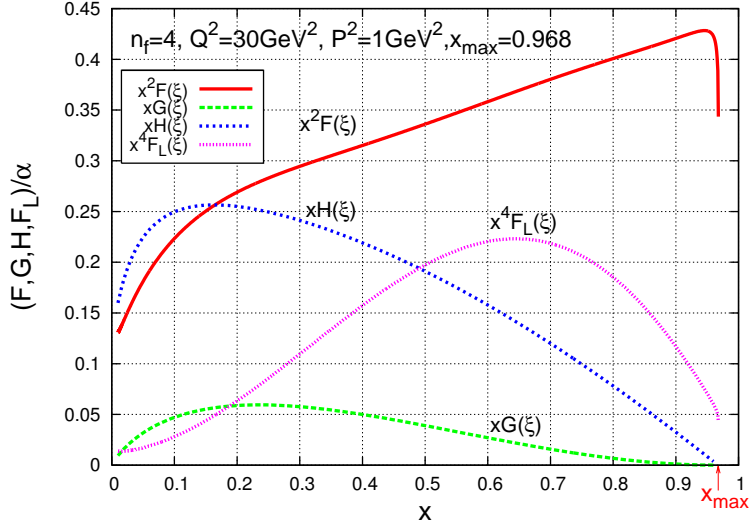}
\vspace{-0.5cm}
\caption{\label{FGHFL} 
The four functions $F$, $G$, $H$ and $F_L$
as functions of $x$. $Q^2=30{\rm GeV}^2$ and
$P^2=1{\rm GeV}^2$. $x_{\rm max}=0.968$.
}
\end{center}
\end{figure}

\begin{figure}
\begin{center}
\includegraphics[scale=1.0]{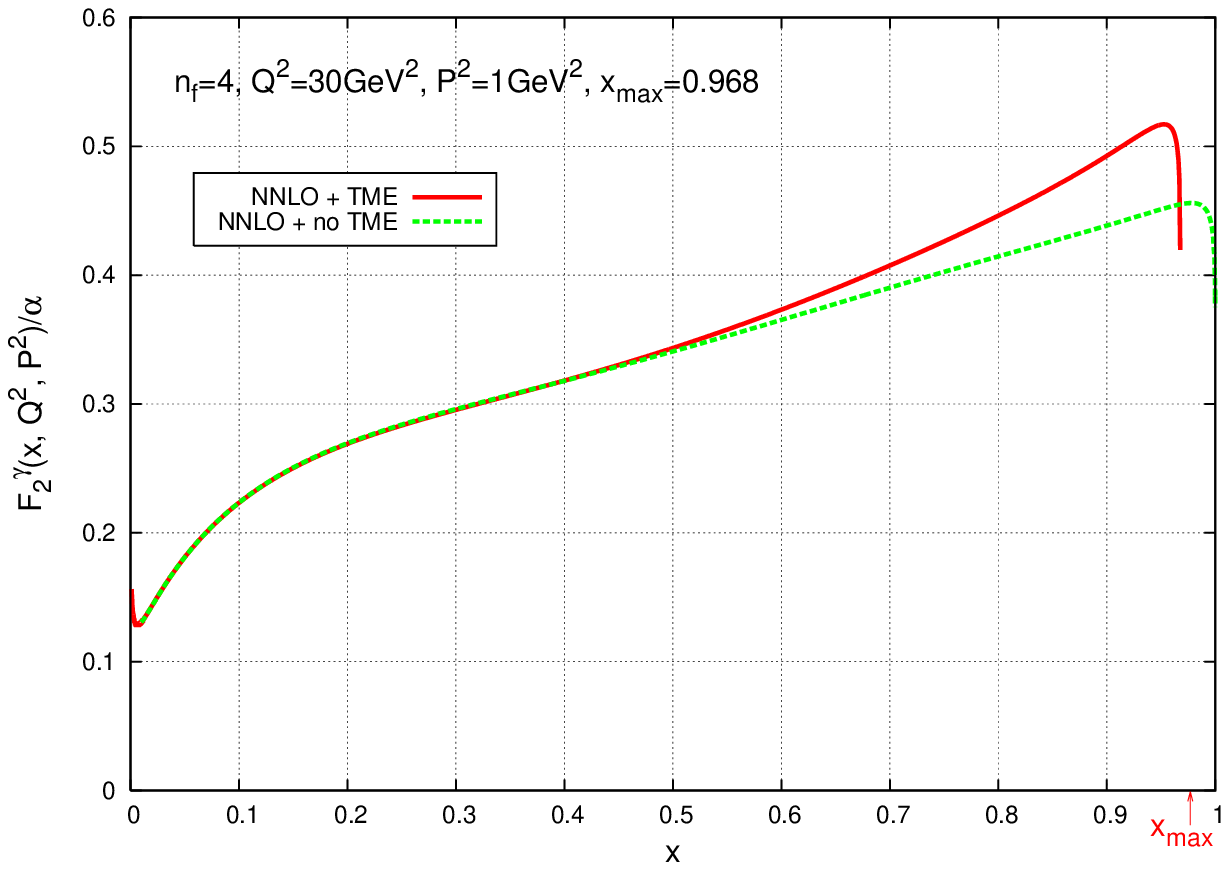}
\vspace{-0.5cm}
\caption{\label{F2} 
$F_2^\gamma(x,Q^2,P^2)$ as a function of $x$ for 
$Q^2=30{\rm GeV}^2$ and $P^2=1{\rm GeV}^2$ with 
$x_{\rm max}=0.968$.
}
\end{center}
\end{figure}

\begin{figure}
\begin{center}
\includegraphics[scale=1.0]{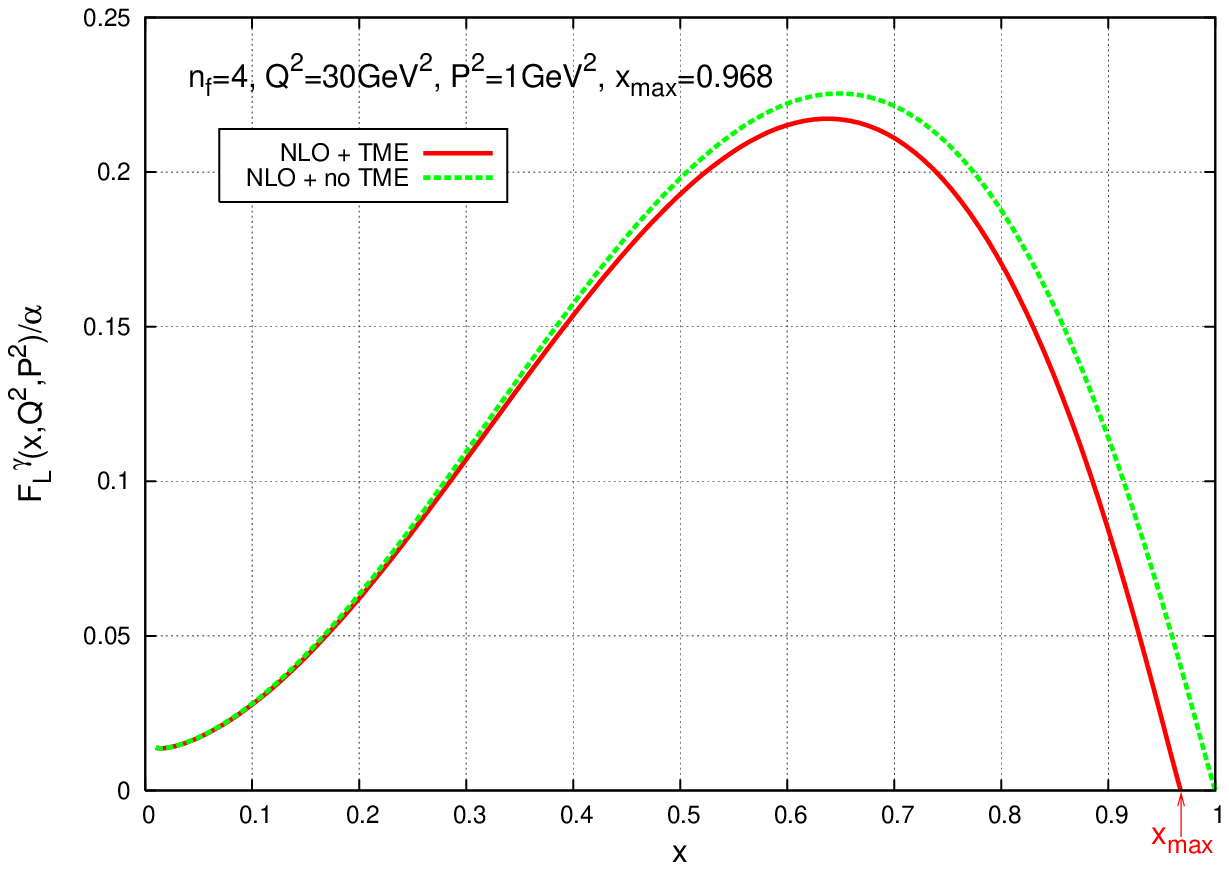}
\vspace{-0.5cm}
\caption{\label{FL} 
$F_L^\gamma(x,Q^2,P^2)$ as a function of $x$ for 
$Q^2=30{\rm GeV}^2$ and $P^2=1{\rm GeV}^2$ with 
$x_{\rm max}=0.968$.
}
\end{center}
\end{figure}

One should note that  $F(\xi)$ is dominant at larger $x$ region, and determines the leading behavior of $F_2^\gamma$. And the factor $x^2/r^3$ in front of
$F_2^\gamma$ shows the deviation upwards from $x^2$ as $x$ approaches 
$x_{\rm max}$. 
On the other hand, the expression for $F_L^\gamma$, Eq.(\ref{TargetMassFL}), 
possesses no dependence upon the function $F(\xi)$, in 
contrast to the case of $F_2^\gamma$. This is a reason why $F_L^\gamma$ 
becomes maximum in the middle $x$ region, as seen from Fig.3. 

\begin{figure}
\begin{center}
\includegraphics[scale=1.0]{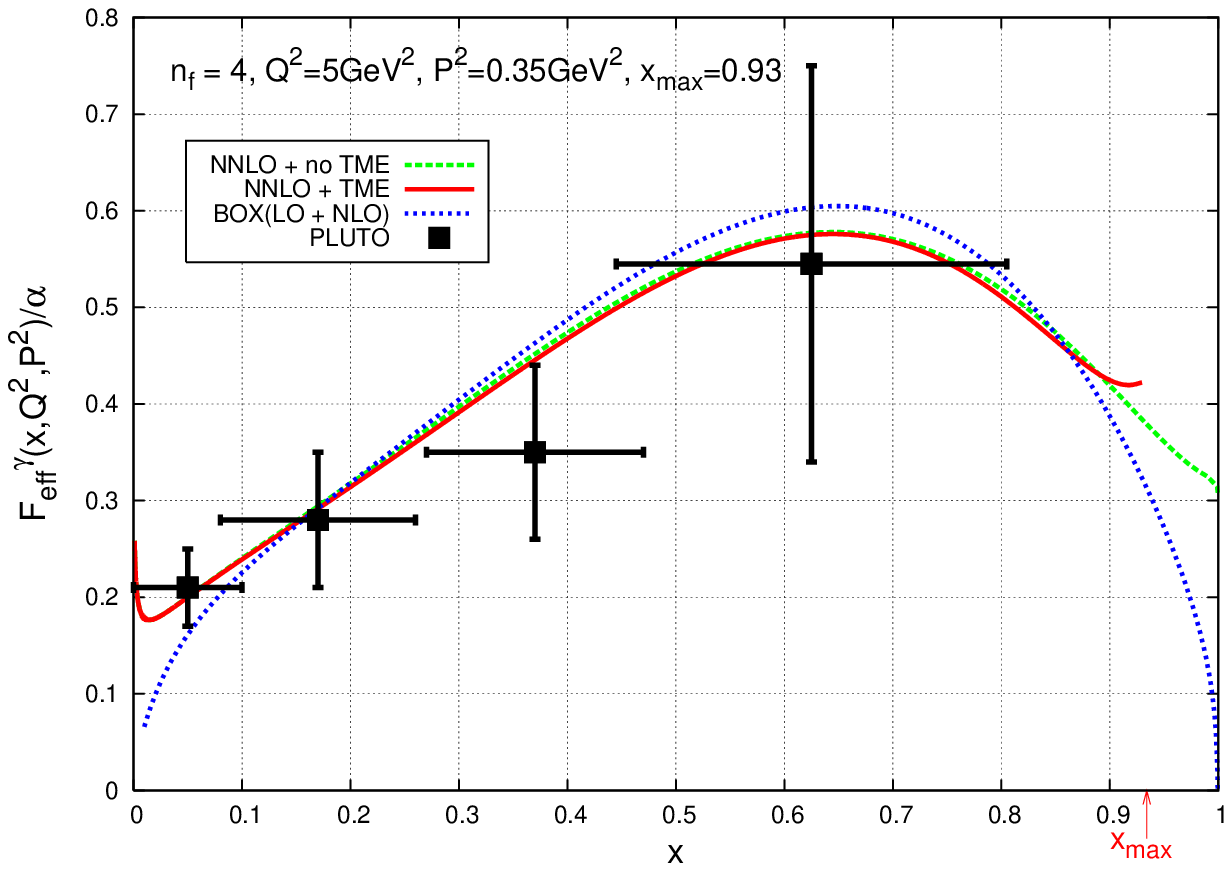}
\vspace{-0.5cm}
\caption{\label{Feff1}
NNLO predictions with and without TME for 
the effective photon structure function:
$F_{\rm eff}^\gamma=F_2^\gamma+\frac{3}{2}F_L^\gamma$ for 
$Q^2=5{\rm GeV}^2$ and $P^2=0.35{\rm GeV}^2$ with 
$x_{\rm max}=0.93$. 
The experimental data are from the PLUTO group \cite{PLUTO}.
The Box diagram prediction to the NLO order is also shown (blue short-dotted
line).
}
\end{center}
\end{figure}

\begin{figure}
\begin{center}
\includegraphics[scale=1.0]{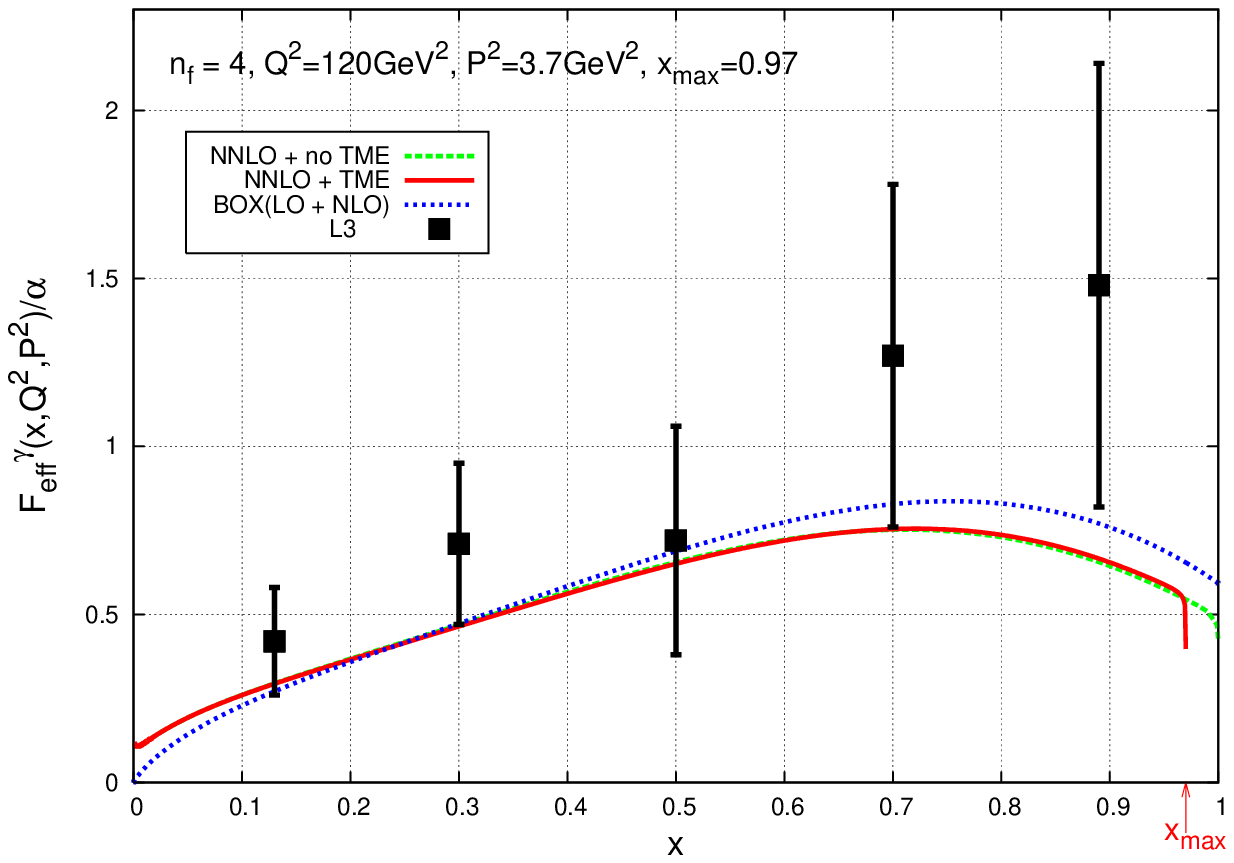}
\vspace{-0.5cm}
\caption{\label{Feff2} 
NNLO predictions with and without TME and Box prediction (NLO)
on 
$F_{\rm eff}^\gamma=F_2^\gamma+\frac{3}{2}F_L^\gamma$ for 
$Q^2=120{\rm GeV}^2$ and $P^2=3.7{\rm GeV}^2$ with 
$x_{\rm max}=0.970$. The data are from the L3 Group \cite{L3}.
}
\end{center}
\end{figure}

\vspace{10cm}
Now let us compare our theoretical prediction for the virtual photon structure
functions with the existing experimental data.
In Fig.6 and Fig.7, we have plotted the experimental data from PLUTO 
Collaboration \cite{PLUTO} and also those from L3 Collaboration \cite{L3} 
on the so-called \lq\lq effective photon structure function\rq\rq defined 
as $F_{\rm eff}^\gamma=F_2^\gamma+\frac{3}{2}F_L^\gamma$, together
with the theoretical predictions. The effective structure function
is proportional to  $\sigma_{TT}+\sigma_{LT}+\sigma_{TL}+\sigma_{LL}$, where 
$\sigma_{ab}$ ($a,b=T,L$; $T=$ transverse and $L=$ longitudinal) is the 
total cross section with the helicity state (a) of the probe photon and 
helicity (b) of the target photon. This combination of $F_2^\gamma$ 
and $F_L^\gamma$ is obtained in the limit of $P^2/Q^2 \ll 1$ 
\cite{GRS,Nisi,GRSch}. 

In the above experiments we have
$Q^2=5\ (120)\ {\rm GeV}^2$ and $P^2=0.35\ (3.7)\ {\rm GeV}^2$ with 
$x_{\rm max}=0.93\ (0.97)$, for PLUTO (L3) data. Here we note that
for the PLUTO
data, $P^2\ll Q^2$ is satisfied since $P^2/Q^2\simeq 0.07$, but
$P^2$ is not much larger than $\Lambda^2$ i.e. $\Lambda^2/P^2\simeq 0.114$.
For L3 data, both hierarchical conditions are satisfied.
Although the experimental error bars are rather large, the data
are considered to be roughly consistent with the theoretical expectations,
except for the larger $x$ region in the case of L3 data. 
Note that the TME for the $F_{\rm eff}^\gamma$ 
in this kinematical region is almost negligible for both cases. 
This could be explained as a consequence of the cancellation of the
TME between $F_2^\gamma$ and $F_L^\gamma$, as discussed above.

\section{Conclusions}

We have investigated the target mass corrections for the unpolarized 
virtual photon structure functions $F_2^\gamma(x,Q^2,P^2)$ and 
$F_L^\gamma(x,Q^2,P^2)$ to the NNLO in perturbative QCD.

In contrast to the case of the nucleon target, the virtual photon
target provides us with the unique testing ground for the perturbatively
calculable target mass effects. 
Taking into account the trace terms in the operator matrix elements 
by using the expansion in terms of orthogonal Gegenbauer polynomials
we get the Nachtmann moments. These moments were then inverted 
to derive explicit expressions for $F_2^\gamma(x,Q^2,P^2)$ and $F_L^\gamma
(x,Q^2,P^2)$ in terms of the four profile functions which are calculable
to NNLO for the first three functions $F$, $G$, $H$ and to NLO for the
last one $F_L$.  
The TME becomes sizable at larger $x$ region, and it enlarges
$F_2^\gamma$ near $x_{\rm max}$ and reduces $F_L^\gamma$
in the region of $x$ larger than the middle point. When we go to 
higher values of $P^2$, e.g. $P^2=3{\rm GeV}^2$ for $Q^2=30{\rm GeV}^2$,
it has turned out that the $F_2^\gamma$ blows up as $x$ approaches
$x_{\rm max}$. So some prescription like resummation of large logs
would be needed to avoid such difficulties. 

We have carried out the confrontation of our theoretical predictions with 
the existing experimental data on the effective photon structure function
$F_{\rm eff}^\gamma$ from PLUTO and also those from
L3 Collaboration. Roughly speaking 
we find the rather good agreement between theory and 
experiments. However, it turned out that TME looks almost negligible 
for $F_{\rm eff}^\gamma$, which is the combination of $F_2^\gamma$ and 
$F_L^\gamma$ and exhibits a cancellation of TME between them.
In the present analysis, we have treated the active flavors as massless
quarks, and ignored the mass effects of the heavy flavors, which should
remain as a future subject. We should also investigate the power corrections
$(P^2/Q^2)^k$ ($k=1,2,\cdots$) due to the higher-twist effects.

We expect the future experiments would provides us with more accurate
data for the double-tag two-photon processes in $e^+ e^-$ collisions.

\begin{acknowledgments}
One of the authors (T.~U.)
would like to thank Guido Altarelli and Silvano Simula
for the useful discussions.
This research is supported in part by Grant-in-Aid
for Scientific Research  from the Ministry of Education, Culture, Sports, Science and Technology, Japan No.18540267. 
\end{acknowledgments}

\newpage
\appendix

\section{\label{StructureFunction}Photon Structure functions}

Averaging the structure tensor $W_{\mu\nu\rho\tau}(p,q)$ given in 
Eq.(\ref{StructureTensor})
over the target polarization, we get~\cite{BCG,SSU} 
\bea
W_{\mu\nu}^\gamma(p,q)&=&-\frac{g^{\rho\tau}}{2}W_{\mu\nu\rho\tau}(p,q)\nn\\
&=&R_{\mu\nu}\Bigl[W_{TT}-\frac{1}{2}W_{TL}\Bigr]
+k_{1{\mu}}k_{1{\nu}}\Bigl[W_{LT}-\frac{1}{2}W_{LL}\Bigr]~, \label{WmunuAppendix} 
\eea
where 
\bea
R_{\mu\nu}&=&-g_{\mu\nu}+\frac{1}{X}\left[p\cdot q
\left(q_{\mu}p_{\nu}+p_{\mu}q_{\nu}  
\right) -q^2p_{\mu}p_{\nu}-p^2q_{\mu}q_{\nu} 
\right]  \\
k_{1\mu}&=&\sqrt{\frac{-q^2}{X}}\left( p_{\mu}-\frac{p\cdot q}{q^2}q_{\mu}  \right)
\eea
with $X=(p\cdot q)^2-q^2p^2$. In the above equation
the first index ($a=T,L$) of the invariant functions $W_{ab}$ refers 
to the probe photon and the second one ($b=T,L$) to the target photon, 
and the subscripts $T$ and $L$ denote the transverse and longitudinal
photon, respectively.

 We define the unpolarized 
photon structure functions $F_2^\gamma$ and $F_L^\gamma$ as,
\bea
\frac{1}{x}F_2^\gamma&=&\frac{(p\cdot
q)^2}{X}\left\{\Bigl[W_{TT}-\frac{1}{2}W_{TL}\Bigr]+
\Bigl[W_{LT}-\frac{1}{2}W_{LL}\Bigr]\right\}~, \\
\frac{1}{x}F_L^\gamma&=&W_{LT}-\frac{1}{2}W_{LL}~.
\eea
Another structure function $F_1^\gamma$ is often used, which is defined 
as~\cite{BergerWagner}
\bea
F_1^\gamma=W_{TT}-\frac{1}{2}W_{TL}.
\eea
Then we get a well-known relation
\bea
F_L^\gamma=-xF_1^\gamma+\frac{X}{(p\cdot q)^2}F_2^\gamma
=-xF_1^\gamma+\left(1-\frac{4x^2P^2}{Q^2}\right)F_2^\gamma~.
\eea
Since $R_{\mu\nu}$ and $k_{1{\mu}}k_{1{\nu}}$ are expressed in terms of 
$e_{\mu\nu}$ and $d_{\mu\nu}$, which are given in Eqs.(\ref{emunu}) and
(\ref{dmunu}), as
\bea
R_{\mu\nu}&=&\frac{p^2q^2}{X}e_{\mu\nu}+\frac{(p\cdot q)^2}{X}d_{\mu\nu}~,\\
k_{1{\mu}}k_{1{\nu}}&=&\frac{(p\cdot q)^2}{X}\{ e_{\mu\nu}+d_{\mu\nu} \}~,
\eea
we find that $W_{\mu\nu}^\gamma(p,q)$ in (\ref{WmunuAppendix}) is rewritten as 
\bea
W_{\mu\nu}^\gamma(p,q)&=&e_{\mu\nu}\left\{\frac{1}{x}
F_L^\gamma+\frac{p^2q^2}{(p\cdot q)^2}\frac{1}{x}F_2^\gamma
\right\}
+d_{\mu\nu}\frac{1}{x}F_2^\gamma~, 
\eea

\section{\label{Gegenbauer}Gegenbauer polynomials}

The Gegenbauer polynomials 
are defined through the generating function given by \cite{Bateman,GRAD}
\bea
(1-2\eta t+t^2)^{-\nu}=\sum_{n=0}^\infty C_n^{(\nu)}(\eta)t^n.
\eea
In terms of  hypergeometric functions $F(\alpha,\beta,\gamma;\, z)$, ~
$C_n^{(\nu)}(\eta)$ is expressed as 
\bea
C_n^{(\nu)}(\eta)&=&
\frac{2^n\Gamma(n+\nu)}{n!\Gamma(\nu)}\eta^n
F\left(-\frac{n}{2},\frac{1-n}{2},1-n-\nu;\frac{1}{\eta^2}
\right) \nonumber\\
&=&\frac{1}{\Gamma(\nu)}\sum_{j=0}^{n/2}
\frac{(-1)^j\Gamma(\nu+n-j)}{j!(n-2j)!}\, (2\, \eta)^{n-2j}~.
\eea 
For example, we have 
\bea
C_n^{(1)}(\eta)&=&\sum_{j=0}^{n/2} \frac{(-1)^j}{j!}\frac{(n-j)!}{(n-2j)!}
(2\eta)^{n-2j}~,  \label{Cn1}
\eea

\subsection{Orthogonality relations}
The orthogonality relation reads
\bea
\int_{-1}^{1}(1-\eta^2)^{\nu-\frac{1}{2}}C_m^{(\nu)}(\eta)C_n^{(\nu)}
(\eta)d\eta
=\frac{2\pi}{2^{2\nu}}\frac{\Gamma(n+2\nu)}{(n+\nu)n![\Gamma(\nu)]^2}
\delta_{mn}.\label{orthogonal}
\eea
In addition we have the following formula for the integral to project 
out the contributions  of definite spin from the dispersion relations,
\bea
&&\int_{-1}^{1}d\eta \ \eta^m
(1-\eta^2)^{\nu-\frac{1}{2}}C_n^{(\nu)}(\eta)\frac{1}
{\zeta-\eta}=\frac{\pi}{2^{\nu-1}}
\zeta^m(\zeta^2-1)^{\frac{\nu-1}{2}}\left[
\zeta-(\zeta^2-1)^{1/2}\right]^{n+\nu}\nonumber\\
&&\hspace{2cm}\times\frac{\Gamma(n+2\nu)}{\Gamma(\nu)\Gamma(n+\nu+1)} 
F\left(\nu,1-\nu,n+\nu+1;\frac{-\zeta+(\zeta^2-1)^{1/2}}{2(\zeta^2-1)^{1/2}}
\right). \label{IntegralGegenbauer}
\eea
In fact the factor $\left[\zeta-(\zeta^2-1)^{1/2}\right]^{n+\nu}$ gives 
$\left(-\frac{P}{Q}\right)^{n+\nu}~\xi^{n+\nu}$, where $\xi$ is the 
so-called $\xi$-scaling variable given in Eq.(\ref{xi}).

\subsection{Recursion relations \label{Recursion}}

The recursion relations for Gegenbauer polynomials
read
\bea
&&nC_n^{(\nu)}(\eta)=2\nu[\eta C_{n-1}^{(\nu+1)}(\eta)-C_{n-2}^{(\nu+1)}(\eta)]~,
\label{R1}\\
&&(n+2\nu)C_n^{(\nu)}(\eta)=2\nu[C_n^{(\nu+1)}(\eta)-
\eta C_{n-1}^{(\nu+1)}(\eta)]~,\label{R2}\\
&&(n+2)C_{n+2}^{(\nu)}(\eta)=2(n+\nu+1)\eta C_{n+1}^{(\nu)}(\eta)
-(n+2\nu)C_n^{(\nu)}(\eta)~.\label{R3}
\eea
We get from (\ref{R1}) and (\ref{R2}),
\bea
C_n^{(\nu)}(\eta)&=&\frac{\nu}{n+\nu}\Bigl[C_n^{(\nu+1)}(\eta)-C_{n-2}^{(\nu+1)}(\eta)
\Bigr]\label{nuPLUS1}\\
C_n^{(\nu)}(\eta)&=&\frac{\nu}{n+\nu}\cdot
\frac{\nu+1}{n+\nu+1}C_n^{(\nu+2)}(\eta)\nonumber\\
&&-\frac{2\nu(\nu+1)}{(n+\nu)^2-1}C_{n-2}^{(\nu+2)}(\eta)
+\frac{\nu}{n+\nu}\cdot\frac{\nu+1}{n+\nu-1}C_{n-4}^{(\nu+2)}(\eta)
\label{Recusion3term}
\eea

Now we derive the second line of Eq.(\ref{tlamp0}).
Choosing $\nu=2$ and $n\rightarrow (n-2)$ in (\ref{nuPLUS1}), we find
\begin{equation}
C_{n-2}^{(2)}(\eta)=\frac{2}{n}\Bigl[C_{n-2}^{(3)}(\eta)-C_{n-4}^{(3)}(\eta)
\Bigr]~.
\end{equation}
Next choosing $\nu=3$ and $n\rightarrow (n-4)$ in (\ref{R3}), we get
\begin{equation}
4\eta C_{n-3}^{(3)}(\eta)
=\frac{2}{n}\left[
(n-2)C_{n-2}^{(3)}(\eta)+(n+2)C_{n-4}^{(3)}(\eta)
\right]~.
\end{equation}
Thus we obtain
\begin{equation}
\Bigl[
C_{n-2}^{(2)}(\eta)-2C_{n-4}^{(3)}(\eta)+4\eta C_{n-3}^{(3)}(\eta)
\Bigr]-2C_{n-2}^{(3)}(\eta)=
\frac{-2}{n}C_{n-2}^{(3)}(\eta)+\frac{2}{n}C_{n-4}^{(3)}(\eta)
\end{equation}

\section{Contraction Formulas \label{Contraction}}
Here we derive Eqs.(\ref{contraction-1}) and (\ref{contraction-2}).
The most general rank-$n$ symmetric and traceless tensor, 
$\{p^{\mu_1}\cdots p^{\mu_n}\}_n$,  that can be formed  with the
momentum $p$ alone,  is expressed as follows~\cite{GP}, 
\begin{equation}
\{p^{\mu_1}\cdots
p^{\mu_n}\}_n=\sum_{j=0}^{n/2}\frac{(-1)^j}{2^j}\frac{(n-j)!}{n!}
\underbrace{g\cdots g}_j \overbrace{p\cdots p}^{n-2j}~(p^2)^j ~, 
\end{equation}
where $\displaystyle{\underbrace{g\cdots g}_j}$ stands for a product of
$j$ metric tensors $g^{\mu_l\mu_k}$ with $2j$ indices chosen among
$\mu_1,\cdots,\mu_{n}$ in all possible ways. Then we easily see that 
the contraction of $\{p^{\mu_1}\cdots p^{\mu_n}\}_n$ 
with $q_{\mu_1}\cdots q_{\mu_{n}}$ is expressed in terms of 
Gegenbauer polynomial $C_n^{(1)}(\eta)$ given in (\ref{Cn1}) as
\cite{NACHT,WAND}, 
\begin{equation}
q_{\mu_1}\cdots q_{\mu_n}\{p^{\mu_1}\cdots p^{\mu_n}\}
=a^n C_n^{(1)}(\eta)~,\label{Izero}
\end{equation}
which is Eq.(\ref{contraction-1}). Here we have put
\begin{equation}
a=-\frac{1}{2}PQ, \quad  \eta=-\frac{p\cdot q}{PQ}\label{variable}~. 
\end{equation}

Next we differentiate both sides of Eq.(\ref{Izero}) twice with respect
to $q_\alpha$ and $q_\beta$. The left-hand side becomes
\bea
\frac{\partial}{\partial q_\alpha}\frac{\partial}{\partial q_\beta}
\Bigl(q_{\mu_1}\cdots q_{\mu_n}\{p^{\mu_1}\cdots p^{\mu_n}\}_n\Bigr)
=n(n-1)q_{\mu_1}\cdots q_{\mu_{n-2}}\{
p^\alpha p^\beta p^{\mu_1}\cdots p^{\mu_{n-2}}\}_n~.  \label{left}
\eea
Differentiation of $a^n C_n^{(1)}(\eta)$ with respect to 
$q_\alpha$ gives 
\bea
&&\frac{\partial}{\partial q_\alpha}\left( a^n C_n^{(1)}(\eta)\right)
=na^{n-1}\left(\frac{-aq^\alpha}{Q^2}\right)C_n^{(1)}(\eta)
+a^n
\cdot 2C_{n-1}^{(2)}(\eta)\left(\frac{p^\alpha}{2a}
+\eta\frac{q^\alpha}{Q^2}\right)\nonumber\\
&&\hspace{3cm}
=\frac{q^\alpha}{Q^2}a^n~ 2 C_{n-2}^{(2)}(\eta)+p^\alpha a^{n-1} C_{n-1}^{(2)}(\eta)
\label{1stDeriv}
\eea
where we have used the following formulas
\bea
\frac{\partial a}{\partial q_\alpha}=a~\Bigl(\frac{-q^\alpha}{Q^2}  \Bigr)~, 
\qquad \frac{\partial \eta}{\partial q_\alpha}=\frac{p^\alpha}{2a}+
\eta\frac{q^\alpha}{Q^2}, \qquad 
\frac{dC_n^{(\nu)}(\eta)}{d\eta}=2\nu C_{n-1}^{(\nu+1)}(\eta)~, \label{DerivETA}
\eea
and,   at the last line,  the recursion relation (\ref{R1}). 
Further, we differentiate the both sides of Eq.(\ref{1stDeriv})
with respect to $q_\beta$~. Again using the formulas in (\ref{DerivETA}) and the recursion
relation (\ref{R1}), we get  
\bea
&&\frac{\partial}{\partial q_\alpha}\frac{\partial}{\partial q_\beta}
\left( a^n C_n^{(1)}(\eta)\right)
=\frac{g^{\alpha\beta}}{Q^2}a^n\cdot 2C_{n-2}^{(2)}(\eta)
+\frac{q^\alpha q^\beta}{Q^4}a^n\cdot
8C_{n-4}^{(3)}(\eta)\nonumber\\
&&\hspace{3cm}+p^\alpha p^\beta a^{n-2}\cdot 2 C_{n-2}^{(3)}(\eta)
+\frac{p^\alpha q^\beta+q^\alpha p^\beta}{Q^2}
a^{n-1}\cdot 4C_{n-3}^{(3)}(\eta)\label{right}
\eea
Thus, from Eqs.(\ref{left}) and (\ref{right}), we obtain
\bea
&&q_{\mu_1}\cdots q_{\mu_{n-2}}\{p^\alpha p^\beta p^{\mu_1}\cdots p^{\mu_{n-2}}
\}_n\nonumber\\
&&\qquad=\frac{1}{n(n-1)}
\left[
\frac{g^{\alpha\beta}}{Q^2}a^n\cdot 2C_{n-2}^{(2)}(\eta)
+\frac{q^\alpha q^\beta}{Q^4}a^n\cdot
8C_{n-4}^{(3)}(\eta)\right.\nonumber\\
&&\qquad\quad \left.+p^\alpha p^\beta a^{n-2}\cdot 2 C_{n-2}^{(3)}(\eta)
+\frac{p^\alpha q^\beta+q^\alpha p^\beta}{Q^2}
a^{n-1}\cdot 4C_{n-3}^{(3)}(\eta)
\right]~,\label{contr}
\eea
which is Eq.(\ref{contraction-2}).
\section{Nachtmann moments of $F_L^\gamma$ \label{FLNachtmann}}
By applying the orthogonality relation (\ref{orthogonal}) for $\nu=3$ 
to the longitudinal amplitude (\ref{tlamp0})
with the help of (\ref{C1C3}) we derive the following recursive relation for 
$M_{L,n}^\gamma$'s: 
\bea
M_{L,n}^\gamma-2\kappa\frac{n+1}{n+4}M_{L,n+2}^\gamma+\kappa^2
\frac{(n+1)(n+2)}{(n+4)(n+5)}M_{L,n+4}^\gamma=I_n\label{MLrecursive}~,
\eea
where
\bea
&&I_n=\int_0^{x_{\rm max}}\frac{dx}{x^5}\xi^{n+3}
\left[\frac{3+3(n+3)r+(n+2)(n+4)r^2}{(n+4)(n+5)}\right]
F_L^\gamma(x,Q^2,P^2)\nonumber\\
&&\hspace{3cm}+4\kappa\frac{1}{n+2}M_{2,n+2}^\gamma
-4\kappa^2\frac{(n+1)(n+2)}{(n+3)(n+4)^2}M_{2,n+4}^\gamma~.
\eea
The above recursive equation (\ref{MLrecursive}) can be solved as an
infinite series:
\bea
M_{L,n}^\gamma=(n+1)\sum_{l=0}^\infty\kappa^l
\frac{(l+1)(n+2+l)}{(n+2l+1)(n+2l+2)}I_{n+2l}~.
\eea
Introducing a variable $t$ defined by ${\xi}(P/Q)\equiv t$ we have
\bea
\xi\sqrt{\kappa}=t,\quad 
r=\frac{1-t^2}{1+t^2},\quad \kappa\xi x=
\frac{t^2}{1+t^2},\quad 
\int_0^{x_{\rm max}}dx\ [\ \cdots]=\int_0^1d\xi\frac{1-t^2}{(1+t^2)^2}
\ [\ \cdots]~,
\eea
and then in terms of $t$ we can sum up the above infinite series and find
\bea
M_{L,n}^\gamma=\int_0^1d\xi\ \xi^n\frac{1-t^2}{1+t^2}
\left[\frac{1}{x^2}F_L^\gamma(x,Q^2,P^2)+4\kappa
\frac{(n+3)-(n+1)t^2}{(n+2)(n+3)}F_2^\gamma(x,Q^2,P^2)\right]~,
\eea
the right-hand side of which turns out to be the Nachtmann moments 
(\ref{NachtmannM_L}),  $\mu_{L,n}^\gamma$.

\end{document}